\documentclass[10pt]{iopart}

\usepackage{color}
\usepackage{cite}
\usepackage{graphicx}% Include figure files
\usepackage{epstopdf}
\usepackage{bm}% bold math
\usepackage{xcolor}
\usepackage{url}
\usepackage[normalem]{ulem}
\usepackage{times}
\usepackage{txfonts}
\usepackage{bbm}
\usepackage{mathrsfs} %Fancy letters 
\usepackage{comment}
\usepackage{dsfont} %added by PJU
\usepackage[colorlinks=true,breaklinks=true,allcolors=blue]{hyperref}
\usepackage{multibib}
\usepackage{soul}

\newcommand{\appropto}{\mathrel{\vcenter{
  \offinterlineskip\halign{\hfil$##$\cr
    \propto\cr\noalign{\kern2pt}\sim\cr\noalign{\kern-2pt}}}}}

\renewcommand{\S}{\textbf{S}}
\newcommand{\rr}{\mathbf{r}}

\newcommand{\innerp}[2]{\langle #1 \vert #2 \rangle}

\newcommand{\xx}{\mathbf{x}}
\begin{document}

\title[The OAM of the atmosphere]{The orbital angular momentum of a turbulent atmosphere and its impact on propagating structured light fields}

\author{Asher Klug$^1$, Isaac Nape$^1$ and Andrew Forbes$^1$}

\address{$^1$School of Physics, University of the Witwatersrand, Private Bag 3, Wits 2050, South Africa}

\ead{andrew.forbes@wits.ac.za}

\vspace{10pt}

\begin{abstract}
When structured light is propagated through the atmosphere, turbulence results in modal scattering and distortions.  An extensively studied example is that of light carrying orbital angular momentum (OAM), where the atmosphere is treated as a phase distortion and numerical tools extract the resulting modal cross-talk.  This approach focuses on the light itself, perturbed by the atmosphere, yet does not easily lend itself to physical insights, and fails to ask a pertinent question: where did the OAM that the beam gained or lost come from? Here, we address this by forgoing the beam and instead calculating the OAM of the atmosphere itself.  With this intuitive model we are able to draw general conclusions on the impact of atmospheric turbulence on OAM beams, which we confirm experimentally.  Our work alters the perspective on this problem, opening new insights into the physics of OAM in turbulence, and is easily extended to other structured light fields through arbitrary aberrations. 
\end{abstract}

\vspace{2pc}

\submitto{\NJP}

\maketitle
 
\ioptwocol
\section{Introduction}

Space division multiplexing and its subsidiary mode division multiplexing are mooted as solutions to the impending data crunch, and seek to exploit the spatial degrees of freedom of light for additional information capacity \cite{Richardson1,Richardson2013A,berdague1982mode,trichili2019communicating}.  This includes using light's spatial modes, tailored in amplitude, phase and polarisation for what is now commonly referred to as structured light \cite{forbes2021structured}.   One popular example is light tailored to have an azimuthally varying phase of the form $\exp(i \ell \phi)$ with integer topological charge (TC) $\ell$ for $\ell \hbar$ of orbital angular momentum (OAM) per photon \cite{allen1992orbital,padgett2017orbital,shen2019optical}.  Although it is understood that other orthogonal mode sets are equally valid, including using the full radial and azimuthal indices of the Laguerre-Gaussian (LG) modes \cite{Trichili2016,zhao2015capacity,zhou2019using} as well as Bessel \cite{mphuthi2018bessel,mphuthi2019free,lukin2014mean,bao2009propagation,zhu2008propagation,nelson2014propagation,ahmed2016mode,cheng2016channel,doster2016laguerre}, Hermite-Gaussian \cite{cox2019hglg,ndagano2017c,Restuccia2016} and vectorial modes \cite{ndagano2017bb,Sit2017,zhu2019compensation,Cox:16}, OAM has become the mode of choice in the majority of communication studies \cite{willner2015optical,Wang2012}.

Unfortunately, structured light becomes distorted in the presence of perturbations, a topical example of which is turbulence in free-space atmospheric links \cite{lavery2017free,krenn2019turbulence,cox2020structured}.  OAM modes in particular experience modal crosstalk, resulting in a spread of OAM. This leads to the deleterious effects of decay in entanglement in quantum states \cite{paterson2005atmospheric,ndagano2017characterizing,zhang2016experimentally,brunner2013,Ibrahim2013,bachmann2019universal,jha2010c,Leonhard2015}, and modal cross-talk for classical communication channels \cite{anguita2008,Anguita2009,Qu2016,Milione2015d,Li2018bb,li2019modeSpaceDiversity,li2020atmospheric,gbur2008vortex}. For this reason, a crucial step towards deploying free space classical\cite{krenn2014communication,krenn2016twisted} and quantum communication channels is the exact characterisation of the turbulence induced perturbations on structured light \cite{steinlechner2017distribution}.

In attempting to investigate the degradation of OAM modes in turbulence, Paterson\cite{paterson2005atmospheric} derived a theoretical expression for the likelihood of scattering between OAM eigenstates, showing that the probabilities were independent of the initial OAM state. Further investigations\cite{tyler2009influence} using optical vortex beams supported these findings with sufficient numerical and experimental confirmation\cite{rodenburg2012influence, malik2012influence}. In contrast, numerical studies of entangled OAM states in turbulence found that higher order single photon LG states were less robust than lower order ones\cite{gopaul2007effect}. Along with the demonstration of a dependence on the initial OAM, other studies have shown that there is also a basis dependence: for example, different results are obtained when using vortex Gaussian beams in comparison to LG, optical vortex \cite{chen2016changes, zhang2020mode} and Ince-Gaussian beams \cite{krenn2019turbulence}. Indeed, while these advancements have elucidated the detrimental effects of turbulence on OAM-carrying light fields, the nuances involved in the modal dependence argument are not settled since there is disagreement with early findings. The uncertainty can in part be attributed to the fact that the focus is on the beam itself, and thus its construction plays a large role in the outcome.

\begin{figure*}[t]
	\centering
	\includegraphics[width=\linewidth]{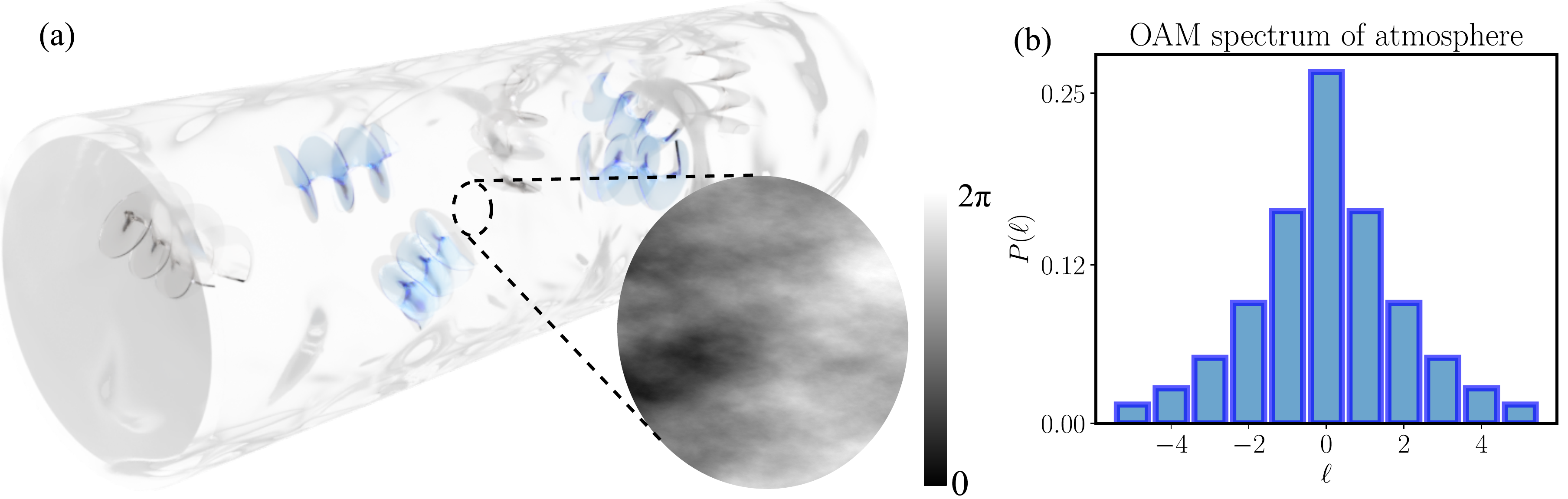}
	\caption{\textbf{Concept of OAM in the atmosphere.} (a) A turbulent `cylinder' of the atmosphere is visualized with spirals, reminiscent of OAM wavefronts, and can modelled as a random phase screen whose frequency spectrum exhibits the -11/3 Kolmogorov power law. (b) The probability of measuring different values of $\ell$ in the atmosphere is illustrated for medium turbulence.}
	\label{fig:concept}
\end{figure*} 

In this work we revisit this topic from an alternative perspective: if the OAM of a beam passing through a turbulent atmosphere changes (gain or loss), then it must have gained or lost the said OAM from the atmosphere itself, since OAM in total (in a closed system) must be conserved.  Therefore, a pertinent question to ask is: what is the OAM of a turbulent atmosphere? How likely is the atmosphere to exchange $\ell \hbar$ of OAM per photon, for some value of $\ell$? Here, we ignore the optical field itself and instead link atmospheric turbulence directly to OAM using Zernike phase aberrations, and show that the atmosphere does have OAM, which it can exchange with optical fields.  By decoupling the field from the atmosphere, we are able to make generalised statements about the impact on structured light beams, with OAM beams a special case.  We use this intuitive model to clear up some confusion in the community on the role of beam size, OAM value and turbulence strength in OAM scattering, and confirm our predictions by laboratory based turbulence experiments.  Our work offers an alternative perspective on the problem, supported by a complete theoretical framework that lends itself to studying arbitrary phase aberrations beyond turbulence in the context of OAM.

\section{Concept and Theoretical Framework}  
In the traditional approach to modelling structured light in the atmosphere, the optical field experiences phase distortions which then results in modal scattering.  In the context of OAM, the beam has some initial OAM (which may be zero or some superposition) with modal scattering amounting to a spread in the OAM values about the initial value(s).  Here, we interpret the spread in OAM as the atmosphere having imparted or removed OAM from the beam.  This allows us to invert our perspective, and ignore the optical field altogether, instead bringing our attention to bear on the atmosphere itself. The question then reduces to: how much OAM is there in a turbulent atmosphere?  Since turbulence is stochastic in nature, we expect the answer to be in probabilities.  

Since we are interested in the atmosphere's ability to transfer OAM with structured light fields, we will quantify the OAM using the discrete spectrum of OAM eigenmodes. We visualize a cylinder of the turbulent atmosphere, as seen in Fig.~\ref{fig:concept}(a). The random pressure and temperature fluctuations in the atmosphere result in density variations that in turn give rise to refractive index fluctuations via the Gladstone-Dale law.  The refractive index variations over the length of the cylinder can be approximated as a thin screen which is described by a function $\Theta$, defined over a radius $R$. We will show how we can find the probability of measuring charge $\ell$ in the atmosphere for a fixed turbulence strength, as seen in Fig.~\ref{fig:concept}(b), from $\Theta$. To this end, we relate OAM to the Zernike polynomials (since they too have an azimuthal term), which are defined on the unit disk by 
\begin{equation}\label{eq:Zernike polynomials}
	C_n^m(\rho,\phi) = \frac{1}{\sqrt{2}}N^{|m|}_n R_n^{|m|}(\rho)\exp\left(im\phi\right),
\end{equation}
with
\begin{equation}\label{eq:normalization for zernike polynomials}
	N^{|m|}_n = 
	\left\{
	\begin{array}{ll}
		\sqrt{2(n+1)}, &\textrm{if } m \neq 0,\\
		\sqrt{n+1}, &\textrm{if } m = 0,
	\end{array} 
	\right.
\end{equation}
and
\begin{equation}\label{eq:radial zernike function}
	R_n^{|m|}(\rho) = \sum\limits_{k=0}^{(n-|m|)/2}\frac{(-1)^k(n-k)!}{k!\left(\frac{n+|m|}{2}-k\right)!\left(\frac{n-|m|}{2}-k\right)!}\rho^{n-2k},
\end{equation}
where $ n,m $ are integers. These (generally) complex-valued polynomials can be rewritten as
\begin{equation}\label{eq:real zernike polynomials}
	C^{\pm|m|}_n(\rho,\phi) = \frac{1}{\sqrt{2}}\left(Z^{|m|}_n(\rho,\phi)\pm iZ^{-|m|}_n(\rho,\phi)\right),
\end{equation}
where
\begin{equation}\label{eq:cosine and sine terms of zernike}
	Z^{\pm|m|}_n(\rho,\phi) = \left\{
	\begin{array}{ll}
		N^{|m|}_n R_n^{|m|}(\rho)\cos\left(|m|\phi\right), &\textrm{for }+|m|,\\
		N^{|m|}_n R_n^{|m|}(\rho)\sin\left(|m|\phi\right), &\textrm{for }-|m|.
	\end{array}
	\right.
\end{equation}
The $ C^m_n $ form a complete set of OAM eigenfunctions, and as such, can be used to write a modal expansion for any well-behaved function. Instead of expanding $ \Theta $ into the $ C^m_n $, we will use the $ Z^m_n $ as they are famously linked to well-known Seidel wavefront aberrations \cite{Born-Wolf-1993}. Using these polynomials will lend greater insight to the behavior of $ \Theta $. Thus we have
\begin{equation}\label{eq:turbulent phase}
	\Theta(r,\phi) = \sum_j\, a_j\,Z_j(\rho,\phi),
\end{equation}
where $ (n,m)\mapsto j $ is an index used for numbering the polynomials. The expansion coefficients are determined in the usual way, with 
\begin{equation}\label{eq:phase expansion coefficients}
	a_j = \innerp{Z_j}{\Theta} = \frac{1}{\pi}\int \textrm{d}^2 \boldsymbol{\rho} \; Z_j(\rho,\phi)\,\Theta(R\rho,\phi),
\end{equation} 
and the integration is performed over a unit disk. 
\begin{figure}
	\centering
	\includegraphics[width=\linewidth]{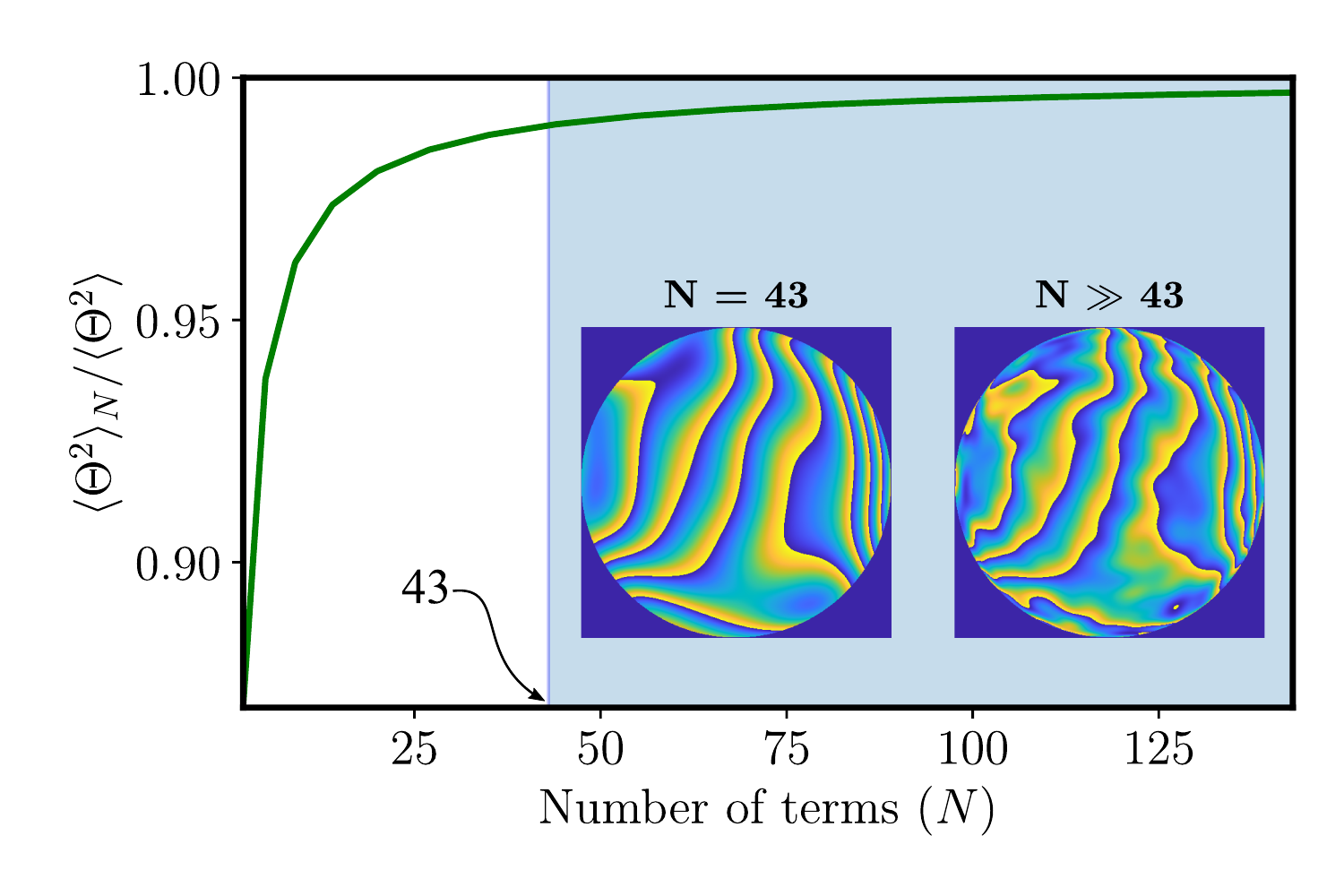}
	\caption{\textbf{Variance of turbulence screen}. A comparison of the variance of a turbulence screen produced from a sum of the $Z_j$ truncated at $N$ terms, $\langle \Theta^2\rangle_N$, to the theoretical variance, $\langle\Theta^2\rangle$. $\langle \Theta^2\rangle_N = 0.99\langle\Theta^2\rangle$ is achieved after 43 terms. Insets show turbulence screens produced from 43 and $\gg43$ terms.}
	\label{fig:fidelity}
\end{figure}
Turbulence is a stochastic process, and this is reflected in the fact that the $ a_j $ are Gaussian random variables with zero mean. The sum of their variance accounts for the variance of the refractive index distortion by \cite{Born-Wolf-1993} 
\begin{equation}\label{eq:wavefront variance}
	\left< \Theta^2\right> = \sum_j \left< |a_j|^2 \right>,
\end{equation}
where $ \langle\cdot\rangle $ denotes the ensemble average and $ \left< \Theta^2\right> $ is the refractive index variance of the atmosphere, up to a constant factor. Few terms are needed to produce a turbulent screen whose variance closely matches the true variance. Fig.~\ref{fig:fidelity} shows the marginal increase in the fidelity of the variance of a screen produced from $N$ Zernike terms, $\left<\Theta^2\right>_N$, in comparison to the theoretical variance in Eq.~\ref{eq:wavefront variance}. The $ \left<|a_j|^2\right> $ can be found analytically for Kolmogorov turbulence and found to be
\begin{equation}\label{eq:ensemble average of zernike coefficients from Noll}
	\left<a_j^*a_{j'}\right> = \mathcal{C}\delta_{m,m'}\sqrt{(n+1)(n'+1)}\left(\frac{D}{r_0}\right)^{5/3}I_{n,n'}
\end{equation}
where $ \mathcal{C} $ is a constant, $ I_{n,n'} $ is the Noll covariance matrix\cite{Noll1976} and $r_0$ is the Fried parameter\cite{fried1966optical}. There is, however, a complication. The transmission function of the atmosphere is not $ \Theta $ but $ \exp\left(i\Theta\right) $ instead. The probability of measuring OAM associated with TC $ \ell $ in the atmosphere, on average, is then 

\begin{equation}\label{eq:ensemble c-ell for atmosphere}
	\left< |c_\ell|^2\right> = \frac{1}{\pi}\int\textrm{d}^2\boldsymbol{\rho}\; \cos\left(\ell\phi\right)\exp\left(-\frac{1}{2}\mathcal{D}\left(\boldsymbol{\rho}\right)\right).
\end{equation}
The function $ \mathcal{D} $ is called the phase structure function, and is given by
\begin{equation}\label{eq:phase structure function}
	\mathcal{D}(\boldsymbol{\rho}) = \sum_j\sum_{j'}\left< a^*_{j}a_{j'}\right>\Delta Z_j(\boldsymbol{\rho})\Delta Z_{j'}(\boldsymbol{\rho}),
\end{equation}
where
\begin{equation}\label{eq:delta zernike}
	\Delta Z_j(\boldsymbol{\rho}) \equiv Z_j(\rho,\phi)-Z_j(\rho,0). 
\end{equation}

\begin{figure}
	\centering
	\includegraphics[width=\linewidth]{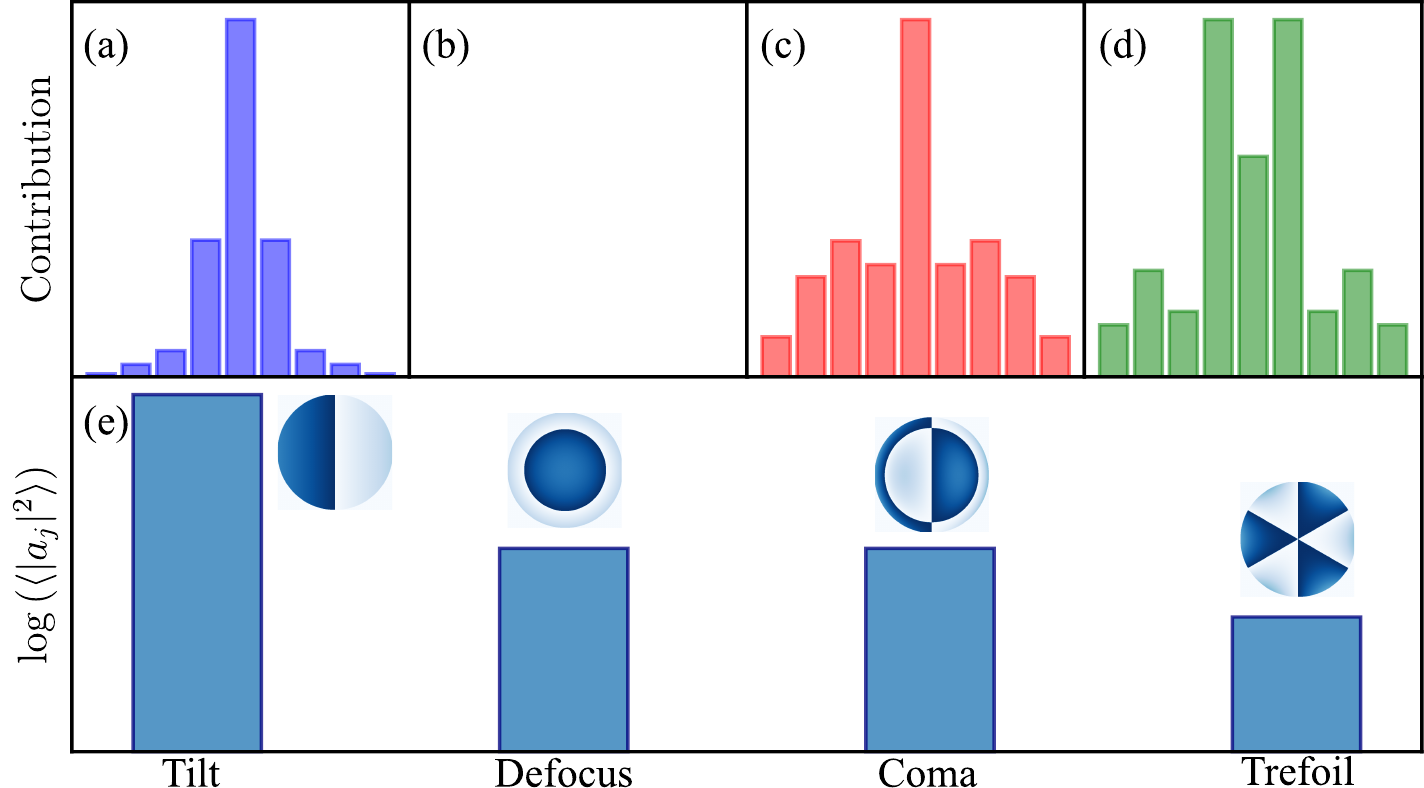}
	\caption{\textbf{Contribution to OAM}. (a) -- (d) Modified OAM spectra of the atmosphere showing the contribution from the various Zernike aberrations normalized to the peak. The aberrations shown, from (a) to (d), are tilt, defocus, coma and trefoil. (e) Logarithmic plot of the variances of the coefficients $a_j$, with insets of the Zernike functions modulo $2\pi$.}
	\label{fig:zernike and oam}
\end{figure}

We can ask: how do the different terms in Eq.~\ref{eq:phase structure function} contribute to the OAM spectrum of the atmosphere? The contribution to the shape of the OAM spectrum, shown in Fig.~\ref{fig:zernike and oam}(a)--(d), is a function of $\Delta Z_j\Delta Z_j'$ alone. The modal spectrum from the tilt terms is centered around $\ell = 0$, whereas that of coma introduces larger $\ell$. The $\left< a_j a_{j'}\right>$ are much larger for lower order terms, as seen in Fig.~\ref{fig:zernike and oam}(e). As a result, higher order terms (like coma and trefoil) will contribute to measuring larger $|\ell|$ values and lower order terms will dominate overall as a result of the relative size of the coefficients.

The probability, on average, of the atmosphere exchanging a photon with $ \ell\hbar $ of OAM in is then $ P(\ell) = \left< |c_\ell|^2\right> $: the atmosphere possesses a spread of OAM, as seen in Fig.~\ref{fig:concept}(b).

\section{The OAM of the atmosphere}
Here we use our approach to make general statements about the OAM of the atmosphere without the need to reference specific beam types.  
\subsection{The phase structure function}
The terms in Eq.~\ref{eq:phase structure function} can be truncated to some finite sum, the limit of which is determined by the required accuracy of the model.  The OAM contribution of higher order terms diminishes rapidly so that for a fixed turbulence strength, the low order terms such as tip and tilt ($ j=2 $ and $ j=3 $) are an order of magnitude greater in coefficient value than the others (Fig. \ref{fig:zernike and oam}). Using just these two terms we can find an analytic expression for the phase structure function, given by 
\begin{equation}\label{eq:Zernike phase structure function}
	\mathcal{D}(\boldsymbol{\rho}) \approx 7.21\left(\frac{D}{r_0}\right)^{5/3}\left(\rho\sin\left(\frac{\phi}{2}\right)\right)^2,
\end{equation}

\noindent where $ D = 2R$. Using this simplified version of our model, we have been able to recover to a high level of accuracy a well-known expression for this function in Kolmogorov turbulence \cite{tyler2009influence}

\begin{equation}\label{eq:exact phase structure function}
	\mathcal{D}(\boldsymbol{\rho}) = 6.88\left(\frac{D}{r_0}\right)^{5/3}\left(\rho\sin\left(\frac{\phi}{2}\right)\right)^{5/3}.
\end{equation}
This approximation further allows us to obtain a closed-form expression for $P(\ell)$, given by
\begin{equation}\label{eq:closed form expression P(ell)}
    P(\ell) \approx \frac{\beta^{\ell}\, _2F_2\left(\frac{1}{2}+\ell,1+\ell;2+\ell,1+2\ell,-2\beta\right)}{2^{\ell} \Gamma\left(2+\ell\right)},
\end{equation}
where $\beta = 1.8025(D/r_0)^{5/3}$, $\Gamma(\cdot)$ is the gamma function and $_2F_2
$ is the generalized hypergeometric function. Take the case $\ell=0$ as an example: Eq.~\ref{eq:closed form expression P(ell)} reduces to
\begin{equation}\label{eq:closed form expression P(0)}
    P(0) \approx  \left(I_0(\beta )+I_1(\beta )\right)\exp\left(-\beta\right),
\end{equation}
where $I_{n}(\cdot)$ is the modified Bessel function of the first kind.
\begin{figure}
    \centering
    \includegraphics[width=\linewidth]{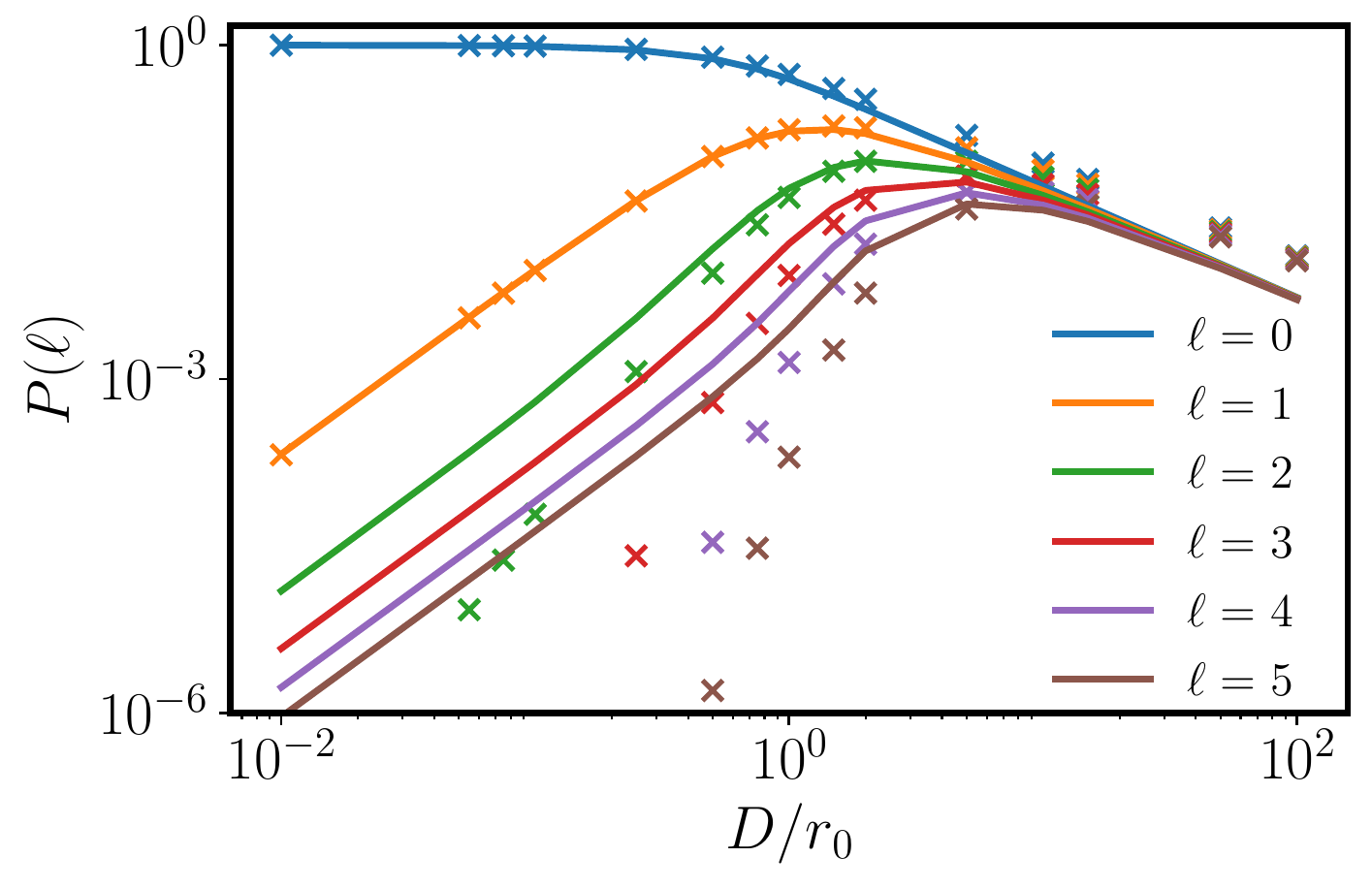}
    \caption{\textbf{Exact and approximate probabilities}. The probability of measuring $\ell = 0$ to 5 in the atmosphere is calculated from both the exact (Eq.~\ref{eq:ensemble c-ell for atmosphere}) and approximate (Eq.~\ref{eq:closed form expression P(ell)}) expressions for weak to strong turbulence. Solid lines and crosses refer to the exact and approximate cases, respectively.}
    \label{fig:P(delta)vsDr}
\end{figure}
Fig.~\ref{fig:P(delta)vsDr} illustrates the comparison between the exact (plotted as solid lines) and approximate (plotted as crosses) probabilities calculated using Eq.~\ref{eq:ensemble c-ell for atmosphere} and Eq.~\ref{eq:closed form expression P(ell)} respectively from weak to strong turbulence. The approximate probabilities are in excellent agreement for the $\ell = 0,\,1$ cases. However, Eq.~\ref{eq:closed form expression P(ell)} underestimates $P(\ell)$ for weak turbulence in general.  

\subsection{The OAM spectrum}

The discrepancy between the refractive index variance and the truncated Zernike sum increases with increasing turbulence strength, so more $ a_j $ are needed. As the $ \left< |a_j|^2 \right> $ are an indication of `how much' OAM is found in $ \Theta $, a more turbulent atmosphere is associated with a greater spread in OAM. Stronger turbulence should therefore result in the presence of higher OAM orders as more $Z_j$ are required for an accurate decomposition.

Further, the spread of OAM in the atmosphere is symmetric about $ \ell=0 $, since the substitution $ \ell\mapsto-\ell $ in Eq.~\ref{eq:ensemble c-ell for atmosphere} leads to
\begin{equation}\label{eq:symmetry of expansion coefficients}
	\left< |c_{-\ell}|^2\right> = \frac{1}{\pi}\int\textrm{d}^2\boldsymbol{\rho}\; \cos\left(-\ell\phi\right)\exp\left(-\frac{1}{2}\mathcal{D}\left(\boldsymbol{\rho}\right)\right) = \left< |c_{\ell}|^2\right>
\end{equation}

\noindent courtesy of the even cosine function. Thus we do not expect a preference for clockwise or anticlockwise helicity in the atmosphere's OAM.  We expect a single peak in the OAM spectrum of the atmosphere around $ \ell = 0 $, as $ \left| \cos(\ell\phi)\right| \leq 1 $, which means that 
\begin{equation}\label{eq:comparison of integrands}
\left| \cos\left(\ell\phi\right)\exp\left(-\frac{1}{2}\mathcal{D}\left(\boldsymbol{\rho}\right)\right)  \right|	\leq \left| \exp\left(-\frac{1}{2}\mathcal{D}\left(\boldsymbol{\rho}\right)\right)  \right|
\end{equation}
over the unit disk, while $P(0) = \left< |c_0|^2\right>$ will have the greatest value. The existence of a second peak is not supported by this theory, in contradiction to some numerical findings \cite{bachmann2019universal}. Since $ \left< \left|c_{-\ell}\right|^2\right> = \left<\left|c_\ell\right|^2\right> $ there can be no $ P(\ell) $ different from $ P(-\ell) $. A second peak would indicate that the spectrum was not symmetric. 

\section{OAM exchange with the atmosphere}
Here, we consider the interaction of the atmosphere with structured light.  
\subsection{Initial mode independence}
We predict that the OAM of the atmosphere is independent of the structured light beam. To see this, consider an optical vortex (V) mode $ W(r/R)\exp\left(im\phi\right) $, where $W(\cdot)$ is an aperture function, equal to 1 when the argument is less than 1 and 0 otherwise. The likelihood of measuring $ \ell\hbar $ of momentum in the final field is, on average,
\begin{equation}\label{eq:modal decomp of PV beam}
	\left< |c^{V}_\ell|^2\right> = \frac{1}{\pi}\int\textrm{d}^2\boldsymbol{\rho}\; \cos\left(\Delta\phi\right)\exp\left(-\frac{1}{2}\mathcal{D}\left(\boldsymbol{\rho}\right)\right) = \left< |c_\Delta|^2\right>,
\end{equation} 
where $ \Delta = \ell-m $ and  $ \left< |c_\Delta|^2\right> $ is the probability of measuring $ \Delta\hbar $ of OAM in the atmosphere (Eq.~\ref{eq:ensemble c-ell for atmosphere}). That is, the OAM spectrum of the final field is identical to that of the atmosphere, except that it is symmetric about $ \ell = m $ or $ \Delta = 0 $. This means that higher order OAM modes should not behave any differently than lower modes while propagating through turbulence of fixed strength. The atmosphere will produce the same change in OAM for any beam. It is thus better to think of $ \Delta $ and not $ \ell $.  
\subsection{Beam types and size}
The previous analysis has not mentioned the size of the field, but turbulence is invariably related to the scale at which we look. This is encoded by the Fried parameter $ r_0 $, which is the average distance over which points in the turbulent phase will be correlated \cite{fried1966optical}. This means that the turbulence strength is characterised by the ratio $ D/r_0 $, where $ D = 2R $ is the diameter of the section of atmosphere we are studying (see Fig.~\ref{fig:concept}). There are thus two ways to adjust the turbulence strength - rescale the `window' through which you are looking or change $ r_0 $. Larger beams propagate over a greater $ D $, which is effectively stronger turbulence. As a result, the definition of beam size is very important. Take Laguerre-Gaussian modes as an illustrative example. These modes, ignoring constants, have the form
\begin{equation}\label{eq:LG mode}
	LG^m_0(\rr) = \left(\frac{\sqrt{2}r}{w_0}\right)^{|m|}\exp\left(-\frac{r^2}{w_0^2}\right)\exp\left(im\phi\right),
\end{equation}
where the radial index is set to 0 and $ w_0 $ is the waist of the embedded Gaussian. The second moment radius is $ \overline{r^2} = w_0\sqrt{1+|m|} $, which gets larger for higher modes. One may, mistakenly, say that higher order modes (with fixed $ w_0 $) are less robust. But, from the perspective of the atmosphere, higher order modes are larger and `see more' of the atmosphere, which is equivalent to passing through stronger turbulence. However, if the waist of the Gaussian factor is set to $ w_0 \rightarrow w_0/\sqrt{1+|\ell|} $, then all modes have the same second moment radius and should behave similarly. The only important factor is the turbulence strength, which is the size of the beam, all else kept equal. 

\section{Experimental Validation}
\begin{figure*}
	\centering
	\includegraphics[width=\textwidth]{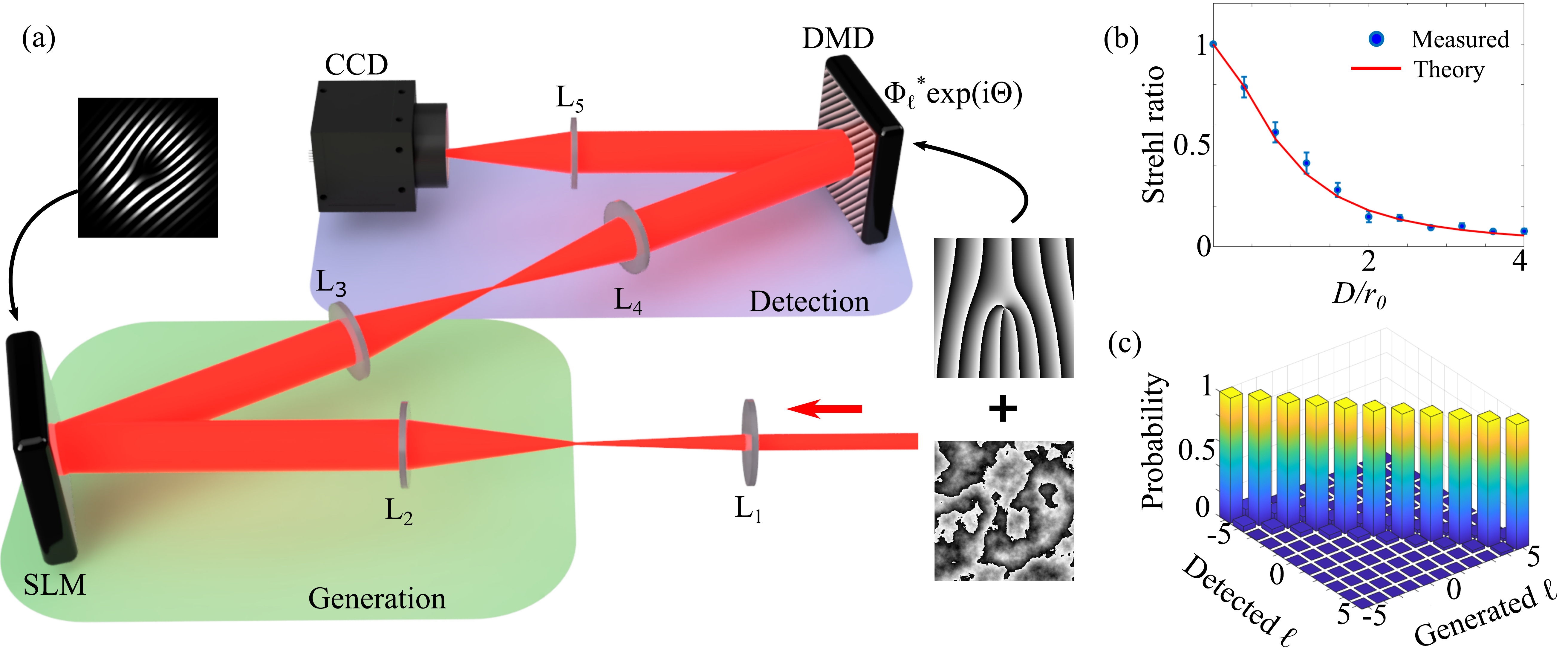}
	\caption{ \textbf{Experimental set-up.} (a)  Lenses $\textrm{L}_1$ and $\textrm{L}_2$ expand a Gaussian beam onto a spatial light modulator (SLM), on which a hologram of a desired initial field is displayed. The field is subsequently imaged by lenses $\textrm{L}_3$ and $\textrm{L}_4$ onto a digital micromirror device (DMD), where binary holograms encoding the detection mode functions, $\Phi_\ell = \exp(i\ell\phi)$, together with the turbulence transmission function, $\exp(i\Theta)$, are displayed (see insets). The final beam is then mapped to the far field by lens $\textrm{L}_5$ where a charge-coupled device (CCD) records an axis intensity corresponding to the modal overlap. (b) The Strehl ratio is plotted for programmed (Theory) and experimentally measured (Measured) turbulence strengths. (c) Crosstalk matrix for generated and detected modes in the range $\ell$ = [-5,5] for zero turbulence.}
	\label{fig:experimentalsetup}
\end{figure*}
\begin{figure}
    \centering
    \includegraphics[width=\linewidth]{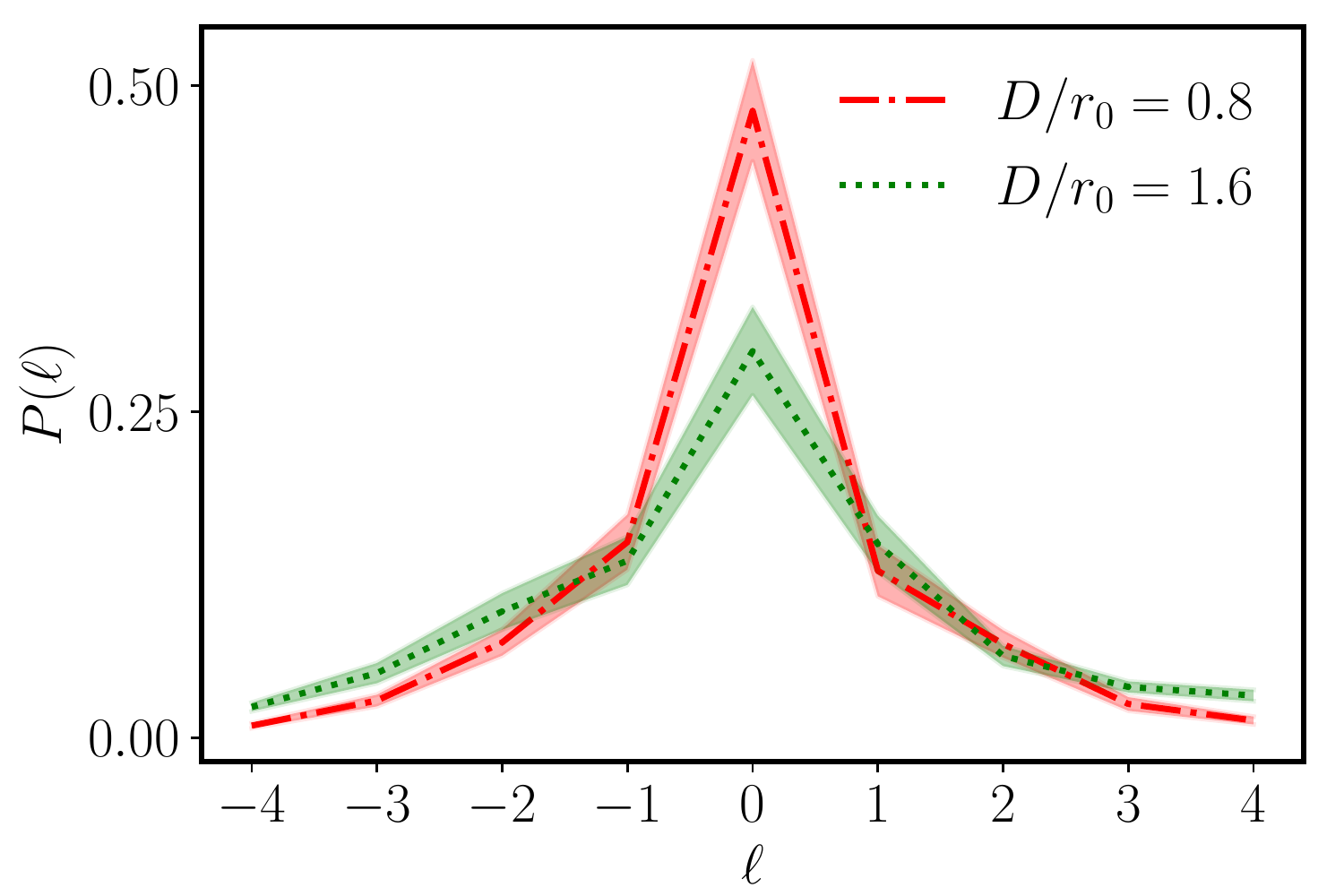}
    \caption{\textbf{Measured OAM of the atmosphere.} Experimental OAM spectrum of the atmosphere for turbulence strengths $D/r_0 = 0.8$ and 1.6. The shaded regions show the variance of $P(\ell)$ around the respective mean values (dashdot and dotted lines). The shaded bands correspond to the measurement uncertainties.}
    \label{fig:oam of atmosphere}
\end{figure}

In this section, we provide experimental confirmation of our theoretical predictions. Our experiment is shown in Fig.~\ref{fig:experimentalsetup}(a) and comprises two parts: a section in which the modes were generated using a Holoeye Pluto phase-only spatial light modulator (SLM), and a second section, where turbulence was added to the mode and the OAM content of the final field was detected using a digital micromirror device (DMD), lens and CCD camera. Lenses $ \textrm{L}_1 $ and $ \textrm{L}_2 $ expanded the laser beam onto the SLM, on which a hologram was displayed. This generated the initial field. Lenses $ \textrm{L}_3 $ and $ \textrm{L}_4 $ imaged the field onto the DMD, on which a hologram of the appropriate match filter and turbulent phase screen was displayed. $ \textrm{L}_5 $ acted as a Fourier lens, ensuring that the far field intensity was observed on the CCD camera where the $P(\ell)$ is proportional to the on axis intensity \cite{pinnell2020modal}. The holograms that were encoded on the SLM and DMD were computed following methods from Refs. \cite{arrizon2007pixelated} and \cite{mirhosseini2013rapid}, respectively. Although the turbulence could be created directly from the Zernike aberrations \cite{Hu1989,Roddier1990,burger2008simulating}, we elected to simulate turbulence by generating phase screens using the sub-harmonic random matrix transform method\cite{lane1992simulation}, which has no direct link to OAM nor the Zernike aberrations. This method guarantees that the screens exhibit the correct $ -11/3 $ power law characteristic of Kolmogorov power spectrum \cite{andrews2005laser}, while ensuring that our predictions are not self-satisfying.

In order to validate the random phase screens we used the Strehl ratio (SR) \cite{cox2020structured}, 
\begin{equation}\label{eq:strehl ratio}
    \textrm{SR} = \frac{\left< I(\mathbf{0})\right>}{I_0(\mathbf{0})} \approx \frac{1}{\left(1+\left(D/r_0\right)^{5/3}\right)^{6/5}},
\end{equation}
as it provides a direct relationship between measured intensities and turbulence strength. It is the ratio of the average on axis intensity of the beam with, $\left< I(\mathbf{0})\right>$, and without, $I_0(\mathbf{0})$, turbulence. Our results, shown in Fig.~\ref{fig:experimentalsetup}(b), show excellent agreement between the measured value of SR (Measured) and the programmed value (Theory). A crosstalk matrix, seen in Fig.~\ref{fig:experimentalsetup}(c), for 11 modes ($\ell = -5\textrm{ to }5$) was obtained for zero turbulence. Such a matrix ensures that any instance of $P(\Delta\neq0)\neq 0$ is attributable to turbulence alone and not experimental error.

A Gaussian beam whose width was large compared to the DMD screen was used to experimentally measure the OAM of the atmosphere, as this beam approximates the constant intensity used in the derivation of Eq. \ref{eq:ensemble c-ell for atmosphere}. Fig. \ref{fig:oam of atmosphere} shows the experimentally measured OAM spectrum of the atmosphere for $D/r_0=0.8$ and $D/r_0=1.6$. The spectra are peaked at $\ell=0$ as expected (see Eq.~\ref{eq:comparison of integrands}) and are symmetric, implying a lack of preference for left or right helicitiy ($\pm \ell$) in the atmosphere. The statistical variance in the $P(\ell)$ is visualized as shaded bands around the data.

\begin{figure}
	\centering
	\includegraphics[width=\linewidth]{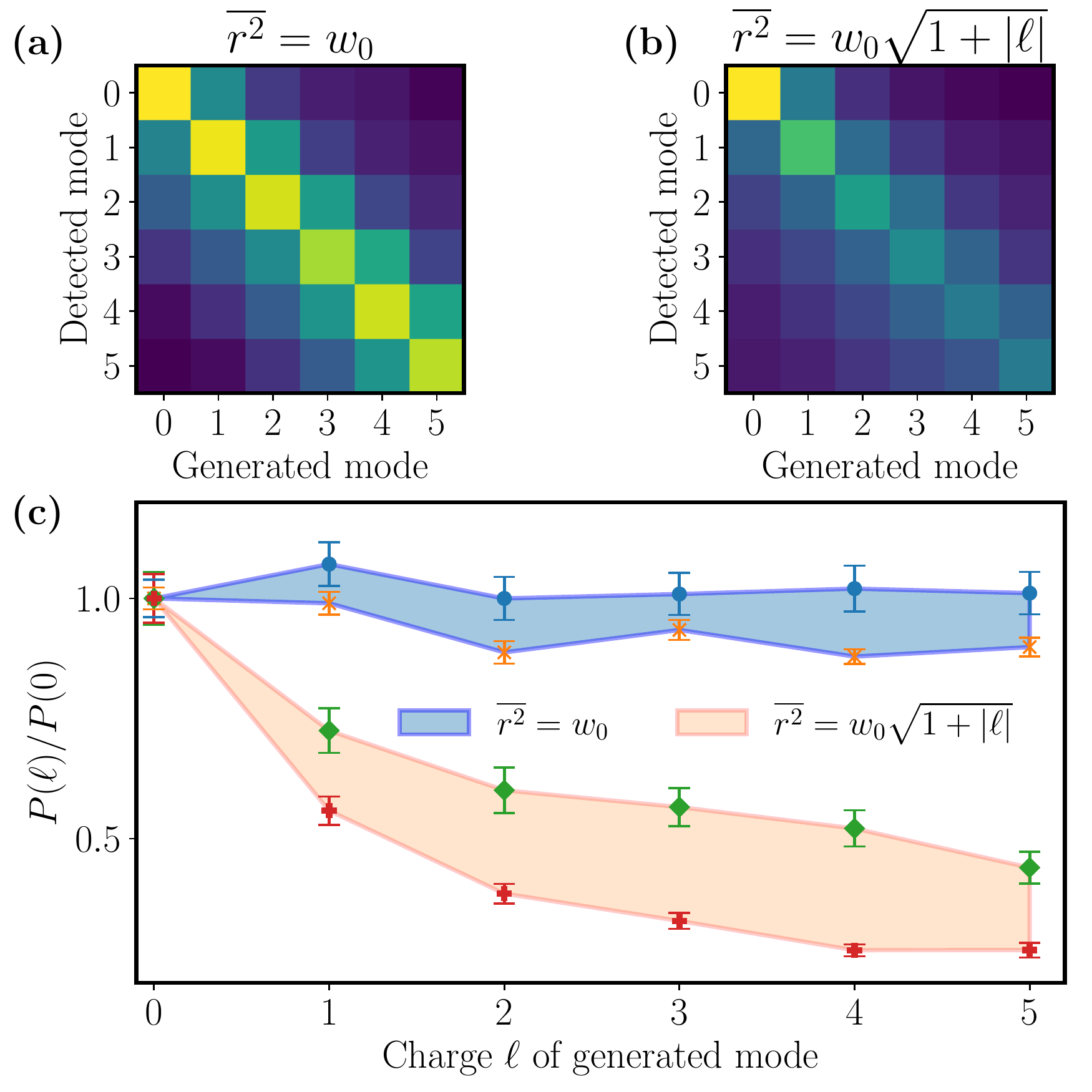}
	\caption{\textbf{Effects of beam resizing on LG modes.} (a) and (b) are crosstalk matrices for LG modes in medium turbulence whose second moment radii are equal to $w_0$ and $w_0\sqrt{1+|\ell|}$ respectively, where $w_0$ is the waist of the Gaussian ($\ell=0$) case.  (c) The quantity $P(\ell)$ is the likelihood of measuring the same charge as the initial beam. The probabilities are normalised to the Gaussian case $P(0)$ for comparison. The upper and lower limits for $\overline{r^2}=w_0$ and $\overline{r^2}=w_0\sqrt{1+|\ell|}$ correspond to $D/r_0=0.8$ and $D/r_0=2.4$, respectively.}
	\label{fig:LGmodes}
\end{figure}

\begin{figure}
	\centering
	\includegraphics[width=\linewidth]{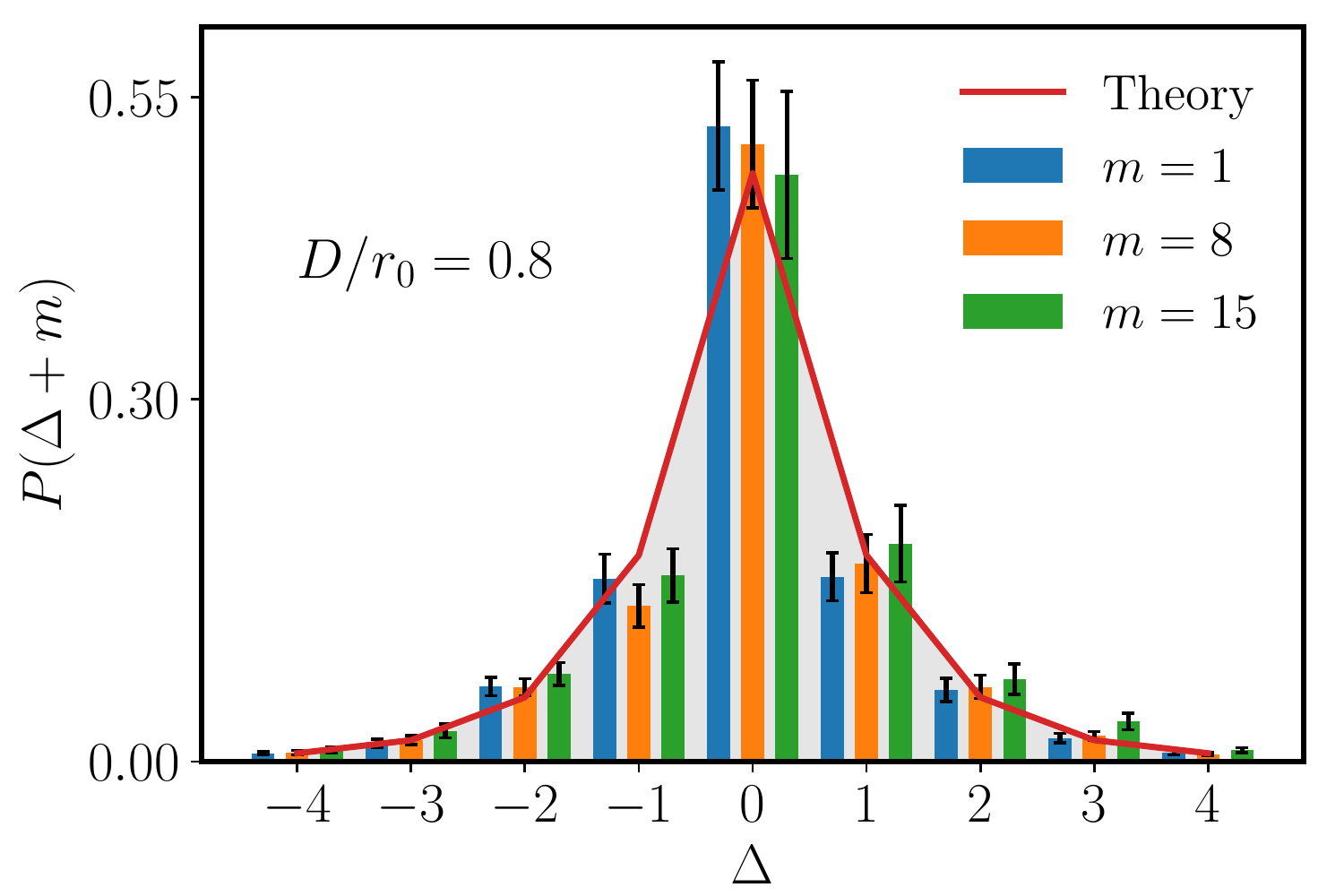}
	\caption{\textbf{Modal independence and symmetry.} The probability of measuring $\ell = \Delta+m$ for PV modes with initial charges $m = 1,\,8$ and 15 in weak turbulence. The solid line corresponds to theoretical values.}
	\label{fig:PVmodes}
\end{figure}

To quantify the interaction between the atmosphere and structured light fields, two families of beams were investigated: vortex LG beams defined in Eq. \ref{eq:LG mode} and pure vortex (PV) modes defined, ignoring constants, as
\begin{equation}\label{eq: PV mode}
	PV_m(\rr) = \exp\left(-\frac{\left(r-r_V\right)^2}{w_0^2}\right)\exp\left(im\phi\right),
\end{equation}
where $ r_V $ is the radial position of the ring. The experiment confirmed the symmetry of the OAM spectra of the beams, the presence of a single peak in those spectra and the independence of initial TC. Fig.~\ref{fig:LGmodes} shows the the dependence of an LG vortex mode's behaviour in turbulence on its size. The crosstalk matrices in Fig.~\ref{fig:LGmodes}(a) and Fig.~\ref{fig:LGmodes}(b) for $D/r_0 = 1.6$ and different beam sizes show how the seemingly unresilient behaviour of higher order modes in turbulence is in fact a result of their increased size. This effect is magnified in Fig.~\ref{fig:LGmodes}(c), where the probability $P(\ell)$ of measuring the same charge as the generated beam (normalized by the Gaussian case $P(0)$) is measured for weak and strong turbulence cases. Beams whose second moment radii are equal perform similarly, independent of the initial OAM charge. In contrast, beams whose size increase with $\ell$ exhibit the well-known mode-dependent degradation as a result of turbulence. PV beams, shown in Fig.~\ref{fig:PVmodes}, were investigated as all modes are the same size. Measurements for modes with initial charges ($ m =1,\,8,\,15$) in weak turbulence are in excellent agreement with the theory: confirming the independence on initial charge and showing the symmetry of the spectrum and the peak at $\ell = 0$, as predicted.
\section{Discussion and Conclusion}

We can clarify the current disagreement in the community regarding the robustness or lack thereof of higher order modes. Studies that supported modal independence \cite{malik2012influence,rodenburg2012influence,tyler2009influence} investigated optical vortex beams and imaged them between the transmitting and receiving planes. These beams lack a mode-dependent size parameter and posses the same second moment radius. In contrast, those studies that reported modal dependence did not resize higher order modes, resulting in larger second moment radii. Propagated vortex Gaussian modes \cite{chen2016changes} will have intensity profiles which depend on the initial charge in a similar way to the more obvious cases of LG \cite{zhang2020mode} and IG \cite{krenn2019turbulence} modes. According to our model, this is all resolved by understanding that turbulence is dependent on a beam's size and not its OAM. When there is a link between the two, e.g. $\overline{r^2}=w_0\sqrt{1+|\ell|}$, one can mistakenly conclude that higher modes are less robust. However, the asymmetry of the OAM spectra \cite{chen2016changes} seems to be an anomalous result. 

The OAM content of the atmosphere is non-zero, even though it possesses a symmetric spectrum which results in a 0 mean charge (refer to Fig.~\ref{fig:oam of atmosphere}). A question may be posed: if the aberrations possess symmetric OAM profiles, which imply a total OAM of 0, how can there be non-zero OAM spreading (modal crosstalk)? This is reconciled by realising that the symmetry of TCs in the aberrations result in a symmetric OAM spectrum. On average, the OAM of the beam is zero. However, there is an equal, non-trivial likelihood that a photon carrying $ (\pm m+\Delta)\hbar = \ell\hbar  $ is present in the final field. Since the beam has gained these photons, the likelihood of measuring photons carrying the initial TC must have decreased. Note that in deriving Eq.~\ref{eq:modal decomp of PV beam}, a rotationally symmetric beam profile is assumed. Although this seems to excludes other families such as the Hermite-Gaussian modes, the same analysis can be applied to these modes by expanding them into a rotationally symmetric basis.

Interestingly, there is a distinction between a Zernike term's contribution to turbulence and its contribution to OAM. This behaviour is reflected in $\mathcal{D}(\boldsymbol{\rho})$ (see Eq.~\ref{eq:phase structure function}). The $\left<|a_j|^2\right>$ are considerably smaller for higher order terms. Thus, turbulence and the OAM spectrum are mostly described by lower order terms. However, Fig.~\ref{fig:zernike and oam} shows the conspicuous zero contribution of the defocus term to OAM, in spite of the relative size of its coefficient. To understand this, consider a beam in a superposition of $\ell=\pm 1$. The phase of such a beam is precisely the inset of the tilt term. Defocus, however, contributes an azimuthally symmetric phase which affects the curvature of the wavefront, but not its vorticity. This is mathematically captured in the result $\Delta Z_4 = 0$, and so defocus results in a null contribution to the phase structure function. 

One may wonder where the OAM of the atmosphere might come from? The notion of the atmosphere possessing angular momentum can be linked to fluid dynamics. Turbulence is characterised by kinetic energy transfer along different length scales. This cascade of energy is facilitated by eddies - regions of non-zero vorticity \cite{antonia1996note}. These regions of circular vortex rings are ultimately responsible for the Kolmogorov -11/3 power law \cite{kiya1991vortex}. These fluid vortices could manifest themselves as phase vortices through the fluctuations in the refractive index of the atmosphere and will be detected as OAM in structured light fields.  We hope that this inspires future research linking these seemingly disparate fields.
\section*{Acknowledgements}
AF acknowledges financial support DSI-CSIR Rental Pool Programme administered by the NRF.

\section*{References}
\bibliographystyle{iopart-num}
\bibliography{main.bib}
\end{document}